# Reliability of stochastic capacity estimates

Igor Mikolasek

Transport Research Centre CDV, Lisenska 33a, 636 00 Brno, Czech Republic

igor.mikolasek@cdv.gov.cz

**Abstract:** Stochastic traffic capacity is used in traffic modelling and control for unidirectional sections of road infrastructure, although some of the estimation methods have recently proved flawed. However, even sound estimation methods require sufficient data. Because breakdowns are rare, the number of recorded breakdowns effectively determines sample size. This is especially relevant for temporary traffic infrastructure, but also for permanent bottlenecks (e.g., on- and off-ramps), where practitioners must know when estimates are reliable enough for control or design decisions. This paper studies this reliability along with the impact of censored data using synthetic data with a known capacity distribution. A corrected maximum-likelihood estimator is applied to varied samples. In total, 360 artificial measurements are created and used to estimate the capacity distribution, and the deviation from the pre-defined distribution is then quantified. Results indicate that at least 50 recorded breakdowns are necessary; 100-200 are the recommended minimum for temporary measurements. Beyond this, further improvements are marginal, with the expected average relative error below 5 %.

## 1  Introduction

Stochastic traffic capacity is an established concept used in traffic modelling and control for one-directional sections of road infrastructure (Brilon et al., 2005; Kianfar & Abdoli, 2021; Lorenz & Elefteriadou, 2001; Shojaat et al., 2018; Wang et al., 2022). It enables realistic modelling of traffic flow (TF) and breakdown behaviour in relevant use cases. It can be especially useful in traffic control applications to establish an operating point that minimizes breakdown probability while maximizing TF intensity, using concepts such as the sustainable flow index (Shojaat et al., 2016) – although it is notably sensitive to the aggregation interval and may not be reliable (Mikolasek, 2025) – or traffic efficiency (Sohrabi & Ermagun, 2018). The corrected maximum likelihood estimator (MLE) should be used to estimate breakdown probability in either case (Mikolasek, 2025).

Multiple methods have been used to estimate the breakdown probability distribution (Arnesen & Hjelkrem, 2018; Brilon et al., 2005; Polus & Pollatschek, 2002). However, some have recently been found to be unsuitable or flawed. Specifically, the Kaplan-Meier estimator (product limit method) is inherently unsuitable for this use case because of the differences between age and lifetime on one side and TF intensity and capacity on the other. Moreover, as hinted above, MLE has been applied incorrectly to this problem in the past (Mikolasek, 2025).

This paper evaluates the reliability of capacity estimates with respect to recorded sample size. It is useful to know how many (reliable) breakdown observations must be recorded before the estimates can be used in modelling or traffic control without the risk of significantly skewing results. This is particularly relevant for temporary bottlenecks such as work zones, where the time frame to gather data is limited.

## 2 Methodology

### 2.1 Maximum likelihood estimation

MLE is a well-established survival analysis method, although it has been misapplied to the capacity problem in the past. Here, it is used to estimate the capacity (breakdown probability) distribution. The correct formulation, derived by Mikolasek (2025), is:

$$\arg\max_{\lambda,\gamma} \ell = \sum_{i=1}^{n}[\delta_i \cdot ln(F_C(I_i)) + (1-\delta_i) \cdot ln(1 - F_C(I_i))], \quad (1)$$

where $\lambda$ and $\gamma$ are parameters of the Weibull distribution (other distributions can be used analogously) optimized by maximizing log-likelihood $\ell$. $\delta_i$ is the failure indicator for observation $i$ (1 – failure/breakdown, 0 – survival, i.e., censored), and $F_C(I_i)$ is the capacity cumulative distribution function (CDF) evaluated at TF intensity $I_i$, which also defines the breakdown probability, since breakdown occurs when current capacity is exceeded. For the Weibull distribution $W(\lambda, \gamma)$, it is:

$$F_C(I_i) = 1 - e^{-(I_i/\lambda)^\gamma}. \quad (2)$$

Refer to Mikolasek (2025) for more details about the concepts of capacity, traffic breakdowns, and censoring.

### 2.2 Synthetic data generation

For empirical data, the true capacity distribution is unknown. To address this, a synthetic data generation approach is employed to evaluate the reliability of the estimates by comparing them against a known, pre-defined distribution. Similar methodologies have been applied in other disciplines, such as geology (Lin & Shearer, 2005) and meteorology (Haupt et al., 2006), when real-world data cannot provide a reliable benchmark due to the absence of ground-truth values.

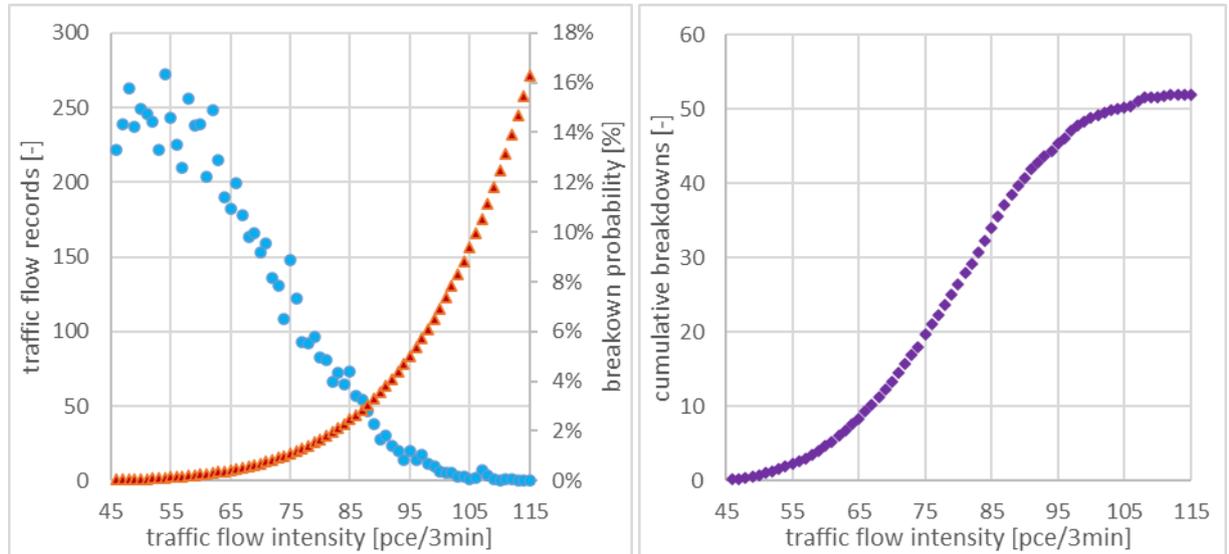

**Figure 1: Left – CDF of capacity representing the breakdown probability ($F_C(I_j)$; red triangles) and the number of records of individual traffic flow intensities ($r_{I_j}$; blue circles). Right – the resulting cumulative frequency of breakdowns calculated via (3) and (4).**

The basis for synthetic data generation was a dataset of real TF intensity records $r_{I_j}$, the number of records at TF intensity $I_j$ (Mikolasek, 2026b) from Mikolasek (2025). Details on data collection and processed are provided there (parameters such as aggregation interval and thresholds may vary by use case). Data from the period without TF harmonisation, comprising 7,447 overlapping 3-minute TF records, 52 of which directly preceded a TF breakdown (exactly one uncensored flow corresponds to one breakdown), were used. While a fully synthetic dataset could be used, employing real data preserves realistic characteristics of the modelled situation and supports the choice of pre-defined capacity-distribution parameters based on the same study. The CDF of capacity can be used to compute the expected, theoretical number of breakdowns $\bar{b}_j$ at each TF intensity level by (3):

$$\bar{b}_j = r_{I_j} \times F_C(I_j). \tag{3}$$

$$CF_B(I_i) = \sum_{j=I_{min}}^{I_i} b_j. \tag{4}$$

The corresponding theoretical (and/or empirical) cumulative frequency of breakdowns ($CF_B$; Figure 1, right) is then calculated via (4), where $b_j$ can be either calculated from (3) (theoretical $CF_B$) or the empirical count. Indices $i$ an $j$ denote different TF intensity levels. The values $\bar{b}_j$ serve as the basis for generating synthetic pseudo-empirical TF breakdown observations $b_j$ using (6). The (pseudo-)empirical $CF_B$ curves do not match the theoretical $CF_B$ exactly, but will fluctuate around it, with deviation depending on the number of recorded breakdowns, which leads to capacity estimation errors.

The synthetic data generation was performed as a series of Bernoulli trials at each level of TF intensity to replicate stochastic variability. For $\bar{b}_j < 1$, the number of breakdowns at level $j$ is modelled as $Bernoulli(\bar{b}_j)$. Implementation is simple: draw $r \sim U(0,1)$ and compare it to $\bar{b}_j$:

$$b_j = \begin{cases} 1 \text{ if } r \leq \bar{b}_j \\ 0 \text{ if } r > \bar{b}_j \end{cases} \tag{5}$$

For $\bar{b}_j \geq 1$, (5) would always yield $b_j = 1$, but $\bar{b}_j$ can be split into $n$ smaller components such that $n \cdot \bar{b}_{j,k} = \bar{b}_j$ and each $\bar{b}_{j,k} < 1$, allowing the Bernoulli trial to be applied separately to each component. This is expressed in (6), where $a_k$ is a realisation of $A_k \sim Bernoulli(\bar{b}_{j,k})$:

$$b_j = \sum_{k=1}^{n}(a_k), a_k = \begin{cases} 1 \text{ if } r \leq \bar{b}_{j,k} \\ 0 \text{ if } r > \bar{b}_{j,k} \end{cases} \tag{6}$$

This yields the number of hypothetical breakdowns $b_j \in \{0, 1, \ldots, n\}$ with expected value $\bar{b}_j$ at each TF intensity level $I_j$. Therefore, $n > \bar{b}_j$ is needed to generate more than the expected number of breakdowns. The resulting pseudo-empirical $CF_B$ curves are then computed via (4).

It is possible to generate virtually an infinite number of synthetic datasets for a hypothetical motorway with a pre-defined capacity distribution. The capacity distribution can then be estimated for each such pseudo-empirical dataset using MLE, and the estimated capacity CDFs can be compared to the pre-defined "true" CDF.

## 2.3 Reliability analysis of the CDF estimates

The ability to generate multiple pseudo-empirical datasets enables a sensitivity analysis of capacity distribution error with respect to sample size. The original dataset with 7,447 TF intensity records yields approximately 52 expected breakdowns under the pre-defined capacity distribution $C \sim W(150; 6.5)$. To assess the effect of sample size, the original TF data were resampled by multiplying (or dividing) the number of records at each level to create eight datasets with 13, 26, 52, 78, 104, 156, 208, and 260 expected breakdowns. Fifteen synthetic pseudo-empirical datasets were generated for each. The underlying capacity distribution remained fixed, ensuring a constant ratio between the expected breakdowns and total TF records (i.e., the censoring rate).

The capacity distribution was then estimated using the corrected MLE formula. For each sample size, the mean and standard deviation of the estimates and the associated errors were computed across the fifteen replications. Note assuming normality to compute the standard deviation is precarious for relative errors, which are bound below at 0 % and have a long right tail; empirical distributions should be considered. The standard deviations are more informative for other variables, such as the estimated Weibull parameters, for which the normality assumption is sounder.

Root mean squared error (RMSE), average relative error (ARE), and average weighted relative error (AWRE; (7)) of the $CF_B$ curves and, primarily, of the capacity CDF were calculated using the pre-defined CDFs and corresponding theoretical $CF_B$ curves as ground truth. Weights $\bar{b}_j$ assign greater weight to TF levels where most breakdowns occur, since larger errors at the outer parts of the estimated curves are typically less consequential for practical performance – assuming traffic patterns and capacity at the site (or at sites with comparable capacity distribution based on similar layout and traffic composition) do not change significantly.

$$AWRE_{CFB/CDF} = \frac{1}{\sum_{i=I_{min}}^{I_{max}} \bar{b}_i} \times \sum_{i=I_{min}}^{I_{max}} \left[ \bar{b}_i \times \frac{(\hat{x}(I_i) - \bar{x}(I_i))}{\bar{x}(I_i)} \right], \text{ where } x \text{ is } CF_B \text{ or } CDF \qquad (7)$$

It was further hypothesised that the censoring rate (determined by the capacity distribution and the demand patterns) affects the reliability of the estimates. To test this, additional simulations were conducted using varied capacity distributions. Specifically, new sample sets were generated by slightly modifying the original theoretical capacity distribution while keeping the expected number of breakdowns approximately aligned with those of the original eight sample sizes again. This was achieved by proportionally multiplying the number of TF records at each level. For eight additional sets, the breakdown probability was reduced by a factor of two ($C \sim W(160; 7.0)$); for other eights sets, by a factor of eight ($C \sim W(183; 7.5)$).

As with the original sets, each new set consisted of 15 synthetic pseudo-empirical datasets, yielding a total of 360 datasets across three censoring levels (capacity distributions), each with eight sample sizes and 15 simulation runs. These were used to estimate a regression model for the average weighted relative error of the estimated $CF_B$ and capacity CDF curves. The total number of TF intensity records (sum of $r_{I_j}$), the number of breakdowns $CF_B(I_{max})$, their ratio (i.e., the censoring rate), and the natural logarithms of these three variables were considered as candidate explanatory variables in the regression model. Model selection was guided by the coefficient of determination $R^2$ and significance of the included variables (P-value < 0.05).

## 3   Results and discussion

Given the extent of the results, only a few illustrative examples of the pseudo-empirical measurements and capacity-model estimation simulations are provided, along with regression models that predict estimation error. The full models and results are available in a data repository (Mikolasek, 2026a).

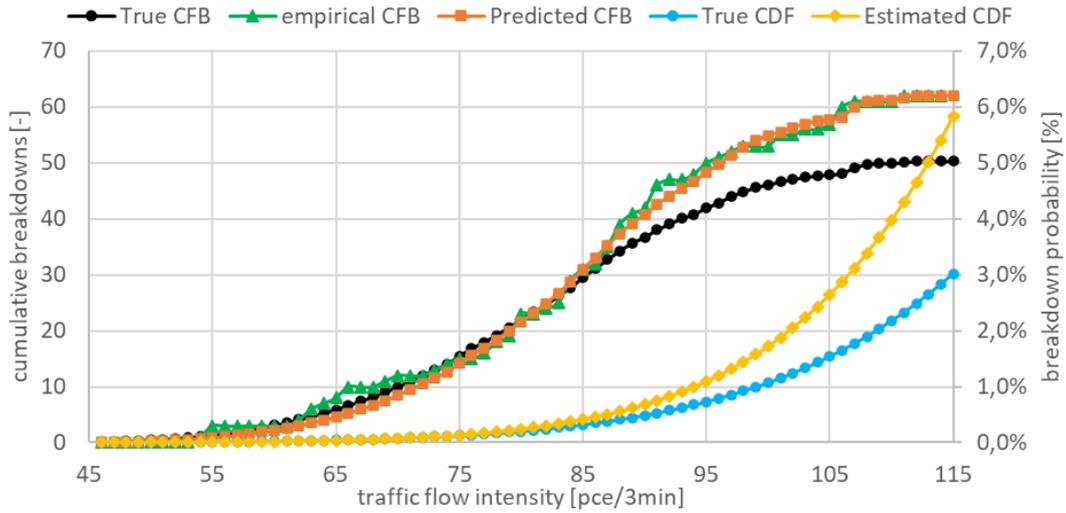

(a)

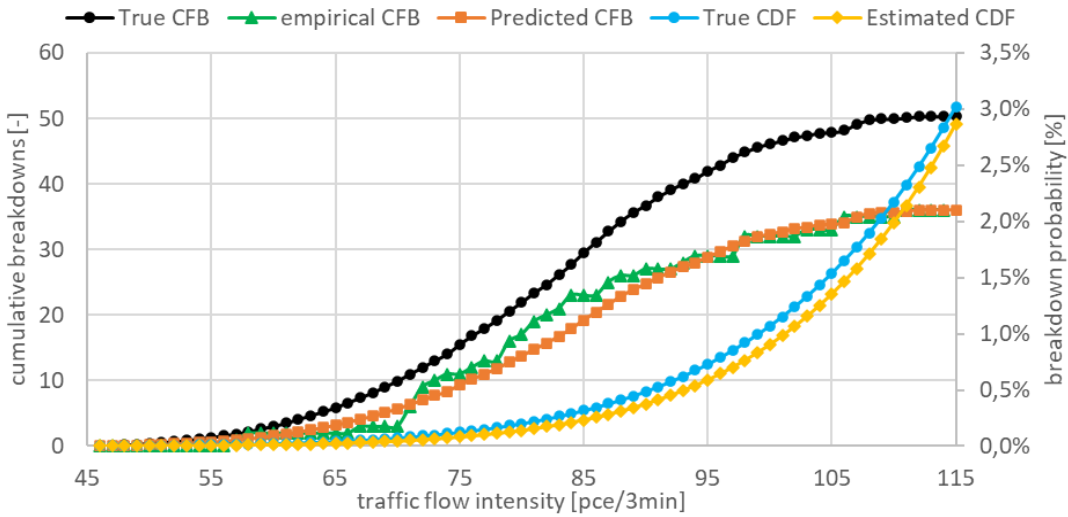

(b)

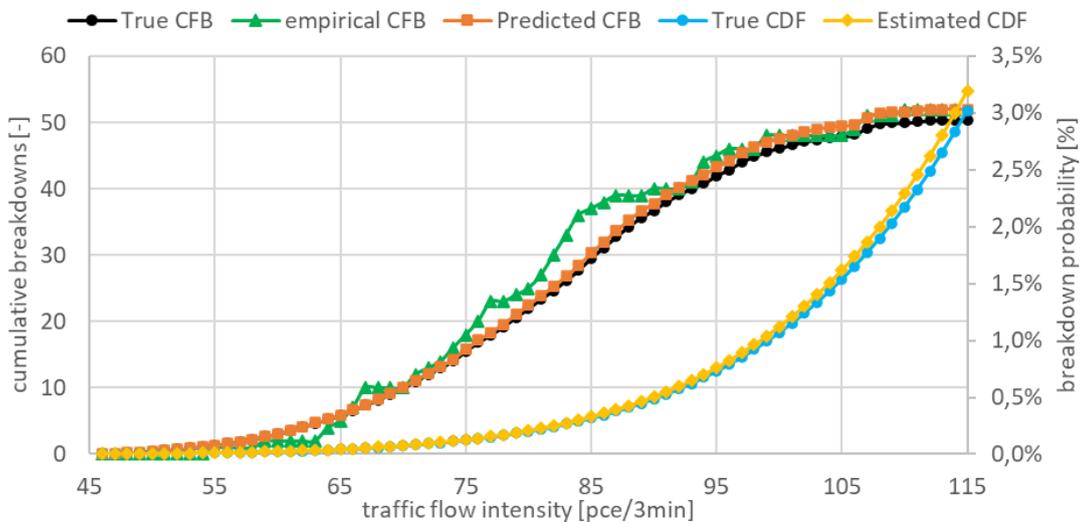

(c)

**Figure 2: Illustrative graphs of the simulated measurements and capacity model estimation from the set with roughly 8x reduced breakdown probability, 8x increased traffic flow data, and 50 expected breakdowns: (a) run 1, (b) run 9, (c) run 11 (see Table 1).**

Figure 2, together with the numbers in Table 1, illustrates the impact of randomness on the accuracy of capacity distribution estimates. Graph (a) shows a smooth empirical $CF_B$ curve that only deviates from the theoretical (unknown in practice) curve mainly in the second half, yet it yields a severely skewed capacity estimate. By contrast, the slightly more rugged curve in graph (c) produces an almost perfect estimate. Graph (b) also shows a relatively smooth $CF_B$ curve that gradually diverges from the theoretical one. The estimated CDF appears to match the true curve visually, but this is a scale effect of the left part of the plot – the average relative error is in fact 29.81% (AWRE = 28.45%).

This highlights that little or no information about reliability of the capacity estimates can be inferred from the $CF_B$ curve shape, tempting as it may be when it is the only available "clue" in practice. While there is a clear correlation between the $CF_B$ and CDF error (Table 1), the $CF_B$ error cannot be calculated, since the theoretical $CF_B$ curve is unknown, and thus cannot be used to infer the CDF error.

Table 1: Excerpt from the simulated capacity estimation results (see Figure 2).

| Run | 1 | 9 | 10 | 11 | 12 | Mean (of 15) | Maximum (of 15) |
|---|---|---|---|---|---|---|---|
| **Records** | | | | 59576 | | | |
| **True shape** | | | | 7.5 | | | |
| **True scale** | | | | 183.0 | | | |
| **Estimated shape** | 8.896 | 8.331 | 7.898 | 7.570 | 8.058 | 7.592 | 9.099 |
| **Estimated scale** | 157.7 | 175.8 | 172.2 | 180.8 | 174.5 | 184.9 | 233.2 |
| **Theoretical breakdowns** | | | | 50.32 | | | |
| **Recorded breakdowns** | 62 | 36 | 59 | 52 | 47 | 49.93 | 62 |
| **Predicted breakdowns** | 62.02 | 36.00 | 59.00 | 52.00 | 47.01 | 49.94 | 62.02 |
| **RMSE CFB** | 5.693 | 9.542 | 4.825 | 0.952 | 2.776 | 4.044 | 9.542 |
| **ARE CFB** | 20.00% | 38.93% | 9.91% | 2.01% | 16.11% | 17.54% | 38.93% |
| **AWRE CFB** | 12.96% | 37.16% | 10.48% | 2.23% | 14.36% | 14.71% | 37.16% |
| **RMSE CDF** | 0.0081 | 0.0011 | 0.0031 | 0.0005 | 0.0010 | 0.0026 | 0.0081 |
| **ARE CDF** | 38.23% | 29.81% | 16.63% | 3.11% | 12.54% | 19.66% | 38.23% |
| **AWRE CDF** | 29.41% | 28.45% | 17.38% | 3.35% | 8.54% | 15.93% | 29.41% |
| **RMSE CFB (vs. emp.)** | 1.604 | 1.427 | 1.070 | 1.783 | 1.214 | 1.494 | 2.827 |
| **ARE CFB (vs. emp.)** | 10.14% | 14.22% | 8.26% | 18.39% | 6.84% | 12.09% | 18.94% |

The calculated errors of the estimated capacity CDF curves cast a new light on earlier studies on stochastic capacity and underscore the need to collect sufficient data before drawing conclusions about the capacity distribution. This primarily concerns the breakdowns, but the intermediate, censored, non-breakdown data must always be included, too. For example, TF harmonisation was recently found to reduce breakdown probability by 40-50 % for a given TF at a 2-to-1 lane motorway work zone (Mikolasek, 2025). Accounting for possible error – up to ± 30 % for either case (with or without the harmonisation), given the number of recorded breakdowns – suggest that, while such a fringe statistical fluke is extremely unlikely, if the possible errors aligned unfavourably, the estimated effect of harmonisation could be in fact negative. However, the impact of harmonisation still appears to be statistically significant; it is also possible that the true effect is even more positive than previously estimated.

Table 2 shows the details of three notable regression models for CDF AWRE. The simple Model 6 was chosen as the most suitable and is shown in Figure 3, which plots CDF AWRE vs. the number of recorded breakdowns and reveals a clear logarithmic relation. While Model 10 has marginally better $R^2$, and Model 4 shows even higher $R^2$, but one variable statistically insignificant (P-value = 0.24), Model 6 is preferred for interpretability and practicality. The censoring rate does not appear to affect estimate reliability. Although it enters Model 10 via the term $\sum r_{I_j}/CF_B(I_{max})$, that term primarily reflects the number of breakdowns; the impact of $\sum r_{I_j}$ is offset by the other (log) term with a negative sign. Table 3 reports analogous models for $CF_B$ AWRE, with the same qualitative conclusions.

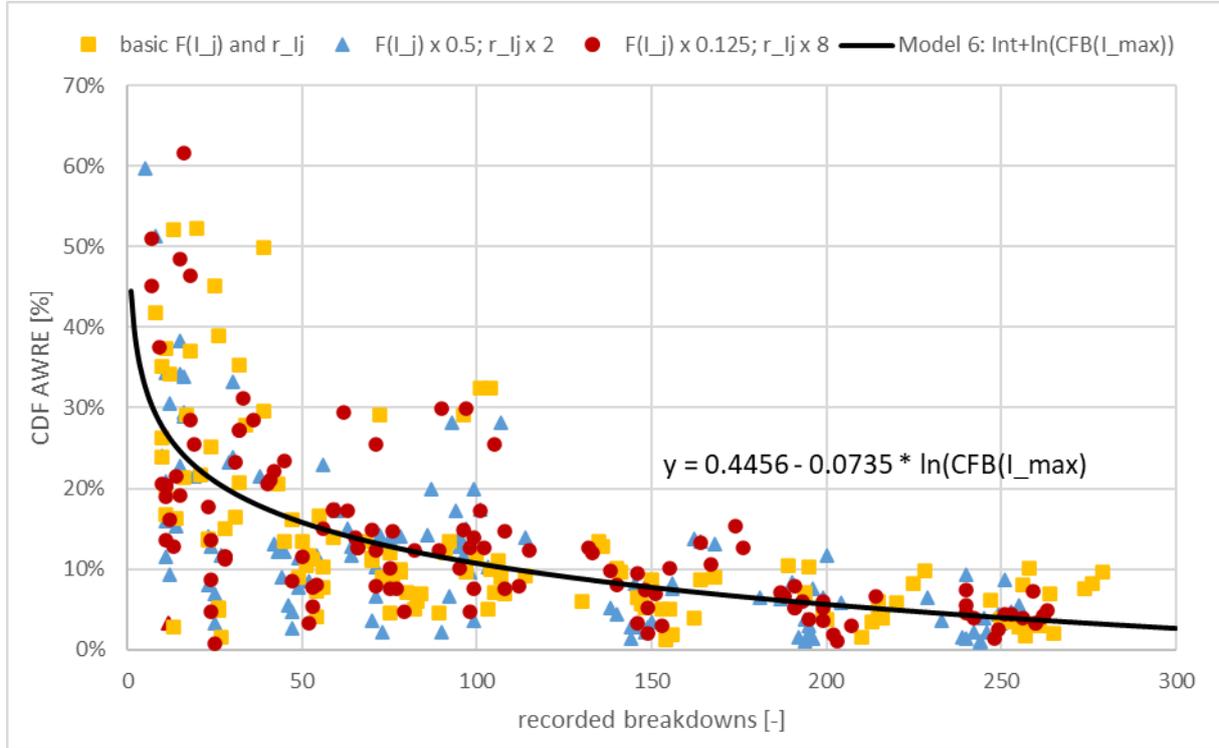

**Figure 3: Regression model of expected breakdown probability (capacity CDF) average weighted relative error (AWRE) and the underlying pseudo-empirical data.**

**Table 2: Notable models for breakdown probability model error (AWRE).**

| Model 4 ($R^2$ = 0.4499) | Coefficients | P-value | Lower 95% | Upper 95% |
| --- | --- | --- | --- | --- |
| Intercept | 0.6854 | 2.671E-12 | 0.4994 | 0.8715 |
| x3: sum $r_{Ij}/CF_B(I_{max})$ | 9.621E-05 | 5.680E-3 | 2.821E-05 | 1.642E-4 |
| x4: ln(sum $r_{Ij}$) | -0.04996 | 8.355E-3 | -0.08701 | -0.01291 |
| x5: ln($CF_B(I_{max})$) | -0.02283 | 0.2427 | -0.06120 | 0.01554 |
| **Model 6 ($R^2$ = 0.4379)** | **Coefficients** | **P-value** | **Lower 95%** | **Upper 95%** |
| Intercept | 0.4456 | 1.882E-72 | 0.4074 | 0.4837 |
| x5 (ln($CF_B(I\_max)$)) | -0.07348 | 1.059E-46 | -0.08213 | -0.06482 |
| **Model 10 ($R^2$ = 0.4477)** | **Coefficients** | **P-value** | **Lower 95%** | **Upper 95%** |
| Intercept | 0.7858 | 4.729E-59 | 0.7074 | 0.8642 |
| x3 (sum r_Ij/$CF_B(I\_max)$) | 1.344E-4 | 1.682E-27 | 1.121E-4 | 1.568E-4 |
| x4 (ln(sum r_Ij)) | -0.07145 | 1.500E-47 | -0.07976 | -0.06314 |

The spread of estimated parameters for the same pre-defined capacity (see Table 1) also sheds a new light fixing the shape parameter as discussed in Mikolasek (2025). Fix it appears even less appropriate, as it limits model's ability to adjust the capacity distribution shape in the relevant region, increasing the risk of severely under- or over- estimating breakdown probability at low or high TF intensities.

Table 3: **Notable models for cumulative frequency of breakdowns model error (AWRE).**

| Model 4 ($R^2$ = 0.3415) | Coefficients | P-value | Lower 95% | Upper 95% |
| --- | --- | --- | --- | --- |
| Intercept | 0.7212 | 1.491E-09 | 0.4928 | 0.9497 |
| x3 (sum r_Ij/CFB(I_max)) | 8.819E-05 | 0.03853 | 4.6807E-06 | 1.717E-4 |
| x4 (ln(sum r_Ij)) | -0.05657 | 0.01497 | -0.1021 | -0.01107 |
| x5 (ln(CFB(I_max))) | -0.01466 | 0.5409 | -0.06178 | 0.03246 |
| **Model 6 ($R^2$ = 0.3282)** | **Coefficients** | **P-value** | **Lower 95%** | **Upper 95%** |
| Intercept | 0.4355 | 2.671E-53 | 0.3887 | 0.4823 |
| x5 (ln(CFB(I_max))) | -0.07141 | 8.663E-33 | -0.08203 | -0.06079 |
| **Model 10 ($R^2$ = 0.3408)** | **Coefficients** | **P-value** | **Lower 95%** | **Upper 95%** |
| Intercept | 0.7857 | 4.140E-44 | 0.6895 | 0.8818 |
| x3 (sum r_Ij/CFB(I_max)) | 1.127E-4 | 9.105E-15 | 8.5354E-05 | 1.401E-4 |
| x4 (ln(sum r_Ij)) | -0.07037 | 3.671E-34 | -0.08055 | -0.06018 |

## 4 Conclusions

It is clearly shown that data volume, specifically the number of recorded breakdowns, plays a crucial role in reliability of estimates of stochastic capacity and the breakdown probability distribution. Based on the results, fewer than 50 breakdowns are likely to yield substantial errors in the estimated capacity distribution. Even with about 100 breakdowns the expected average weighted relative error is roughly 10 % and can commonly reach 30 %. With more than 200 recorded breakdowns, the AWRE is likely to remain below 10 %, with expected value about 5 % or lower as data increase. This is especially important where the traffic flow measurements are temporary, such as in work zones. Where sites are sufficiently similar – or where key variables beyond TF intensity are controlled – data from multiple locations can be pooled to obtain a more robust capacity model. These findings likely generalize to other stochastic capacity estimation methods, though the aggregation interval and TF intensity bin size may affect results and warrant further investigation.

**Acknowledgements**

This article was produced with financial support from the Czech Ministry of Transport within the programme of long-term conceptual development of research institutions.